\documentclass[useAMS,usenatbib]{mn2e}

\usepackage{amssymb, amsmath, epsfig, bm, url, fixltx2e}
\newcommand{\Msun}{{M_{\odot}}}

\newcommand{\HI}{H{\sc ~i}}
\newcommand{\HII}{H{\sc ~ii}}
\newcommand{\HeI}{He{\sc ~i}}
\newcommand{\HeII}{He{\sc ~ii}}
\newcommand{\HeIII}{He{\sc ~iii}}

\begin{document}

\title{On the intergalactic temperature-density relation}
\author[M. McQuinn \& P. Upton Sanderbeck]{Matthew McQuinn$^1$\thanks{mcquinn@uw.edu} and Phoebe R. Upton Sanderbeck$^{1}$\\
$^{1}$Department of Astronomy, University of Washington}


\maketitle\label{firstpage}

\begin{abstract}
Cosmological simulations of the low-density intergalactic medium exhibit a strikingly tight power-law relation between temperature and density that holds over two decades in density.  It is found that this relation should roughly apply $\Delta z \sim 1-2$ after a reionization event, and this limiting behavior has motivated the power-law parameterizations used in most analyses of the Ly$\alpha$ forest.  This relation has been explained by using equations linearized in the baryonic overdensity (which does not address why a tight power-law relation holds over two decades in density) or by equating the photoheating rate with the cooling rate from cosmological expansion (which we show is incorrect).  Previous explanations also did not address why recombination cooling and Compton cooling off of the cosmic microwave background, which are never negligible, do not alter the character of this relation.  We provide an understanding for why a tight power-law relation arises for unshocked gas at all densities for which collisional cooling is unimportant.  We also use our results to comment on (1) how quickly fluctuations in temperature redshift away after reionization processes, (2) how much shock heating occurs in the low-density intergalactic medium, and (3) how the temperatures of collapsing gas parcels evolve.
\end{abstract}


\section{introduction}

\begin{figure}
\begin{center}
\epsfig{file=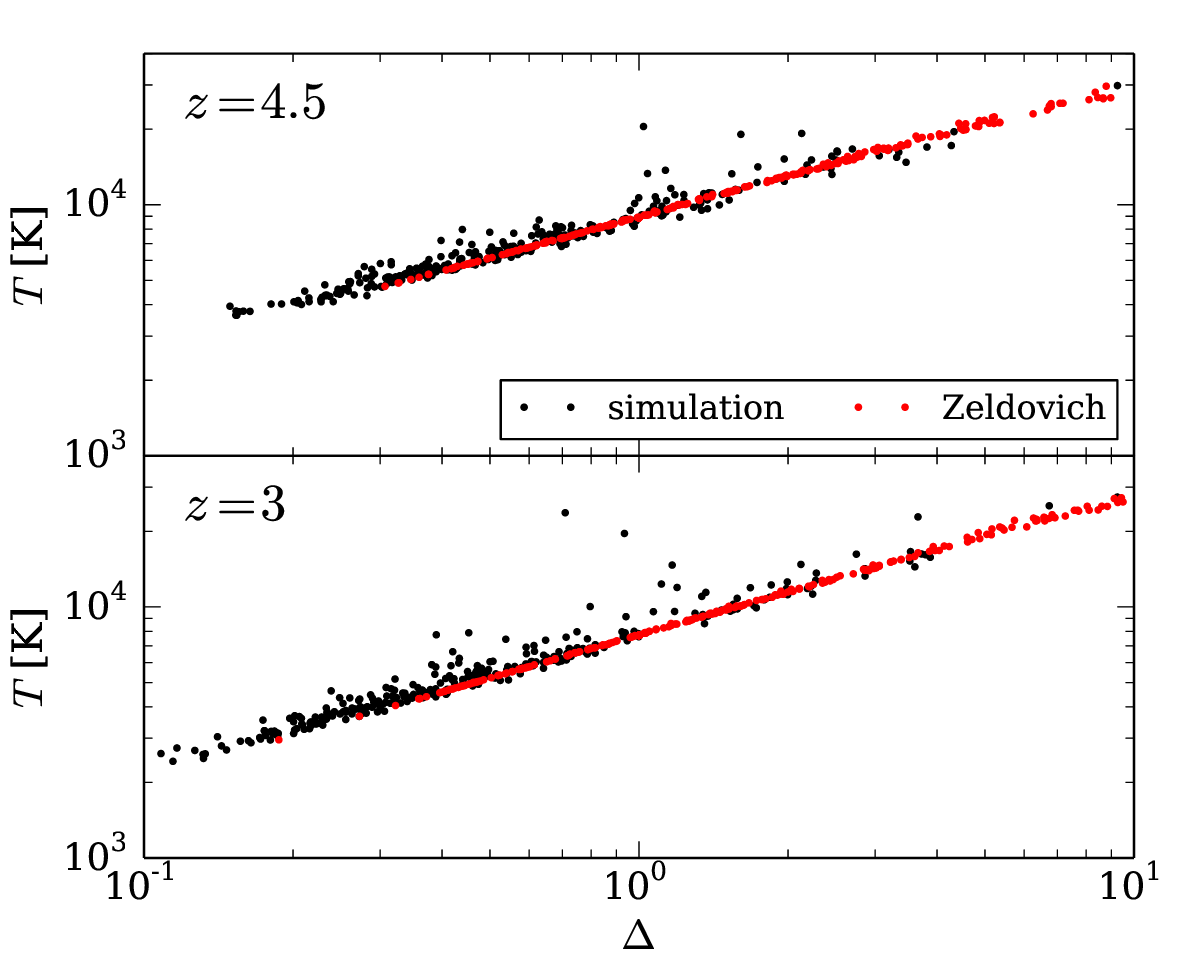, width=8cm}
\end{center}
\caption{Temperature-density relation in a cosmological simulation (showing $300$ randomly selected locations) and in the Zeldovich approximation calculations (following $300$ Lagrangian elements).  Both the simulation and the Zeldovich approximation calculations are initialized with $T_0=20,000~$K and $\gamma-1=0$ at $z=9$, and are evolved forward with $\alpha_{\rm bk} =0$.  The resulting $(T, \Delta)$ values fall on almost a single power-law relation over two decades in $\Delta$, especially in the Zeldovich approximation calculation.  A similar result was found in \citet{hui97}.  The simulation points show more dispersion than the Zeldovich approximation ones because of (mostly weak) shocking (\S\ref{sec:shocks}).
\label{fig:Tdelta}}
\end{figure}

Cosmological simulations of the low-density intergalactic medium exhibit a tight power-law relation between temperature and density that holds over two decades in density (e.g. \citealt{hui97}; see Fig.~\ref{fig:Tdelta}).  This relation is established $\Delta z \sim 1-2$ after the end of cosmological reionization processes \citep{hui97, trac08, furlanetto09, mcquinn09}.
  Motivated by this limiting behavior, a power-law parameterization for the thermal state is adopted in almost all analyses of the Ly$\alpha$ forest (e.g., \citealt{mcdonald05b, viel06, lidz09}).\footnote{Most studies that use numerical simulations implicitly adopt such a parametrization, as standard cosmological simulations employ uniform radiation backgrounds -- an approximation which guarantees a near power-law relation.}  Here we provide a simple explanation for the physics that sets the asymptotic relation as well as an understanding for how quickly this asymptote is reached.  

While the mathematics of what sets the asymptotic temperature and its power-law slope in density were worked out in the seminal study of \citet{hui97}, their linearized approach does not explain why the same power-law holds over a range of $\sim 100$ in density. Subsequently, \citet{theuns98} and \citet{puchwein14} attempted to provide more intuitive derivations that we show are problematic.  In addition, previous studies did not address how non-adiabatic cooling processes affect this relation, namely Compton cooling off of the cosmic microwave background (CMB) and recombination cooling.  Both cooling processes can be many tens of percent of the photoheating rate at the mean density of the Universe (and comparable to photoheating at other relevant densities).  

 We show that the power-law form of the $T-\Delta$ relation occurs because the photoheating rate per $n_b^{2/3}$ happens to be nearly independent of $n_b$, where $n_b$ is the number density of baryons, driving all cosmic gas towards a single adiabat.  This form for the photoheating rate also results in the temperature of a gas parcel being essentially independent of its collapse history -- explaining why the asymptotic $T-\Delta$ relation holds so tightly despite the varied histories of cosmological fluid elements.  (The tightness of this relation is absolutely striking in calculations that do not allow for shocking but allow for varied collapse histories.  See the Zeldovich approximation calculations in Fig.~\ref{fig:Tdelta}.)  We further show that Compton cooling off of the CMB has no effect on the asymptotic slope of $T(\Delta)$ and that recombination cooling is just weak enough not to break the relation.

This paper is organized as follows.  We first enumerate the relevant heating and cooling processes in \S\ref{ss:processes}.  Next, in \S\ref{sec:asymp}, we derive an analytic expression for the asymptotic $T-\Delta$ relation in the limit of only photoheating and adiabatic heating/cooling.  These are the processes considered in previous attempts to understand this relation analytically.   Section~\ref{sec:other_cool} generalizes our solutions to include all relevant cooling processes.  We briefly consider three applications of our results in \S\ref{sec:applications}: (1) how quickly temperature fluctuations are erased, (2) the importance of shock heating, and (3) the thermal evolution of collapsing gas clouds.  Lastly, \S\ref{sec:previous_explanations} contrasts our results with previous attempts to understand the intergalactic $T-\Delta$ relation.   

Throughout we use the standard phenomenological parametrization
\begin{equation}
T = T_0 \Delta^{\gamma-1}
\end{equation}
 for the temperature of the low-density IGM, where $\Delta$ is the baryonic density in units of its cosmic mean, $T_0$ is the gas kinetic temperature at $\Delta=1$, and $\gamma-1$ is a power-law index. (Sometimes $\gamma$ is called the ``equation of state.'')  When necessary, we assume a flat $\Lambda$CDM Universe with $h=0.7$, $\Omega_m = 0.3$, $\Omega_b = 0.045$, and $Y_{\rm He} = 0.25$.  Following \citet{hui97}, we use the Zeldovich approximation to follow the density (necessary for computing the temperature) of fluid elements for certain calculations. Our Zeldovich approximation calculations draw matter elements randomly from the probability distribution of collapse eigenvalues, assuming that the gas traces the dark matter.  This probability distribution only depends on the variance of the density field, which we set to $1.2$.  This choice is equal to the variance of a real space top hat filter containing a Lagrangian mass of $10^{10}~\Msun$ in our reference cosmology at $z=3$ (and approximately the Jeans' mass near the cosmic mean density at this redshift).\footnote{We do not omit regions that have gone through a singularity.  Regions in which an odd number of dimensions have collapsed result in negative densities that are not seen in our plots.  If we were solving the full nonlinear theory, parcels that collapsed along one or more dimensions typically would result in $\Delta \gtrsim 10$, roughly above the densities considered here.}  In addition, we also use a $20~$Mpc, $512^3$ gas and $512^3$ dark matter particle simulation that was run with the GADGET-3 smooth particle hydrodynamic code \citep{springel05}.  Reionization occurs instantaneously in this simulation at $z=9$, heating all gas to $20,000~$K, and the subsequent heating and cooling takes rates applicable for primordial gas if only the \HI\ and \HeI\ are kept ionized with an ultraviolet background that is flat in specific intensity [erg\,s$^{-1}$\,Hz$^{-1}$\,cm$^{-2}$\,sr$^{-1}$].

\begin{table*}
\caption{Heating and cooling rates per baryon for a photoionized IGM, in units of 3860~K per free particle per Gyr.$^*$  The amplitude and power-law indices are evaluated at $T_4=1$, $\Delta = 1$, and $Z_3=1$, where $Z_3 \equiv (1+z)/4$ and $T_4 = T/10^4~$K.  The power-law indices in $Z_3$ and $\Delta$ are exact, and the power law in $T_4$ is accurate over relevant temperatures.}
\begin{footnotesize}
\begin{center}
\begin{tabular}{c c c c c c c c}
\hline \hline
 \HI\ photo. &  \HeI\ photo.  &  \HeII\ photo. & \HII\ recomb. & \HeIII\ recomb. & Compton & Free-free & Hubble\\ \hline
  $ \frac{T_4^{-0.7} Z_3^3\Delta}{1+\alpha_{\rm bk}/2}$  &  $0.13 \frac{T_4^{-0.7} Z_3^3 \Delta}{1+f(\alpha_{\rm bk})}$ & $2.0 \frac{ T_4^{-0.7} Z_3^3\Delta}{1+\alpha_{\rm bk}/2}$   & $-0.11 \,T_4^{0.2} Z_3^3\Delta$    & $-0.20 \,T_4^{0.3} Z_3^3\Delta$  &   $-0.28\, T_4 Z_3^4$ & $-0.05 \,\sqrt{T_4}  Z_3^3\Delta$& $-1.6 \, T_4 Z_3^{3/2}$ \\
\hline
\end{tabular}
\end{center}
\end{footnotesize}
\label{table:simple}\label{table:scalings}
$^*$All rates assume that the applicable bound states have ionization fractions that are either nearly zero or nearly one, showing the choice that maximizes the rate.   We have omitted collisional cooling and \HeII\ recombination cooling, the former is subdominant at $\Delta < 10$ and the latter is always small, with a rate of $-0.011 T_4^{0.3} Z_3^3\Delta$.  The photoheating rates assume photoionization equilibrium, and the normalization of the \HI\ and \HeI\ rates are calculated for a background with $\alpha_{\rm bk} =0$ until an abrupt cutoff at $4\;$Ry.  See footnote~\ref{foot:photo} for additional details such as the definition of $f(\alpha_{\rm bk})$. 
\end{table*}

\section{Relevant heating and cooling processes}
\label{ss:processes}

\begin{figure}
\begin{center}
\epsfig{file=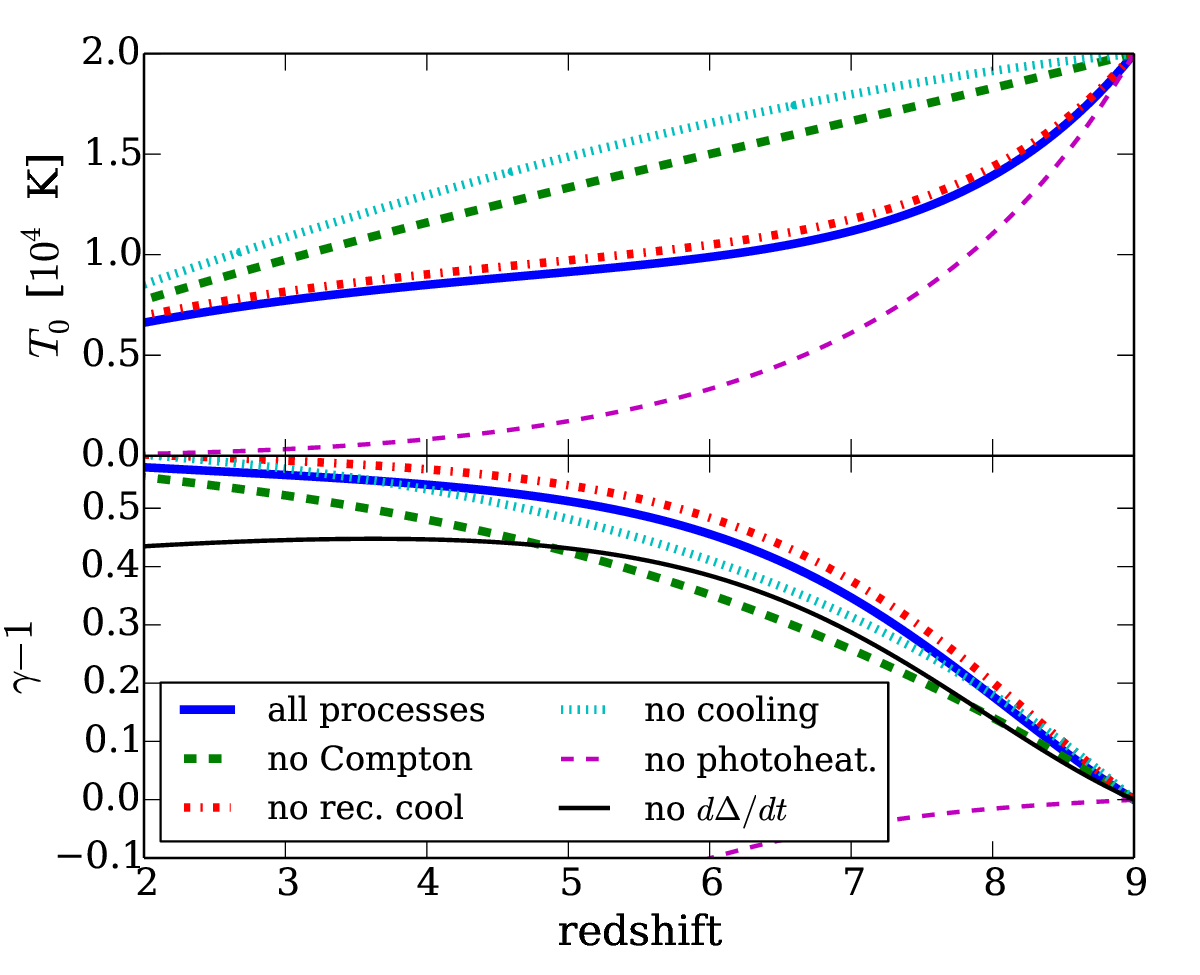, width=8cm}
\end{center}
\caption{Importance of different processes in shaping the temperature at the cosmic mean density, $T_0$ (top panel), and the power-law index of the $T-\Delta$ relation, $\gamma-1$ (bottom panel).  These calculations are initialized with $(T_0, \gamma-1) = (20,000~{\rm K}, 0)$ at $z=9$.  They assume that $\alpha_{\rm bk} =0$ to calculate the photoheating rates and that the \HeII\ has not been reionized such that the gas is comprised primarily of \HII\ and \HeII, with trace amounts of neutrals.  To evaluate $\gamma-1$, we assume that $\Delta = 1$ and that growth  scales so $\Delta - 1 \propto a$. 
\label{fig:T0gamma}}
\end{figure}

\begin{figure}
\begin{center}
\epsfig{file=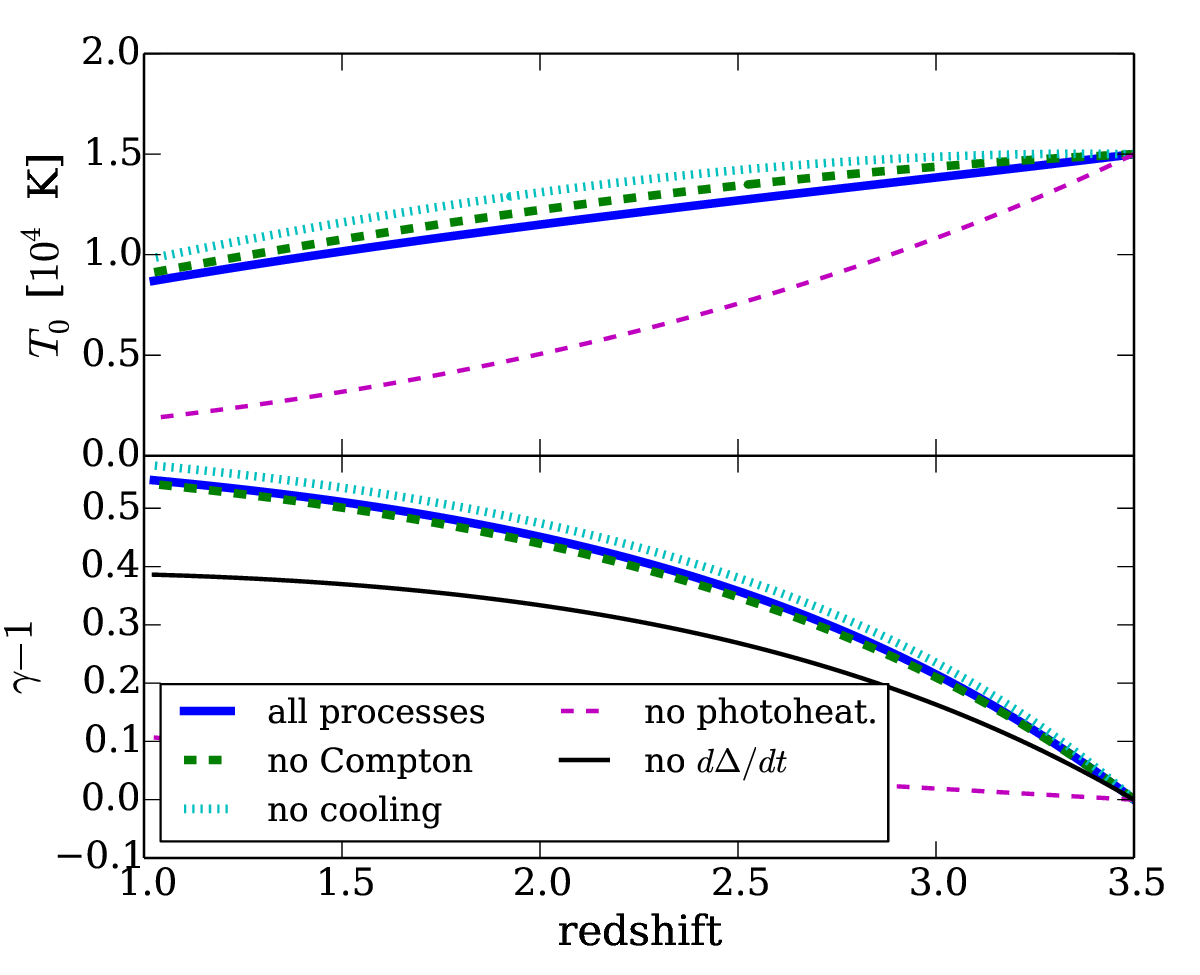, width=8cm}
\end{center}
\caption{Same as Figure~\ref{fig:T0gamma} except for two differences:  First, the calculations are initialized with $(T_0, \gamma-1) = (15,000~{\rm K}, 0)$ at $z=3.5$.  Second, we take photoheating and cooling rates that assume the \HeII\ has been reionized.  
\label{fig:T0gammaHeII}}
\end{figure}

The evolution of the temperature of a Lagrangian fluid element in an FRLW background cosmology follows 
\begin{equation}
\frac{dT}{dt}=-2HT+\frac{2T}{3 \Delta}\frac{d
\Delta}{dt}+\frac{2}{3k_B n_b}\frac{dQ}{dt}, \label{eqn:dTdz}
\end{equation}
where $\Delta$ is the baryonic density in units of the cosmic mean and $n_b$ is the number density of all ``baryonic'' particles in the plasma (including electrons).  Equation~(\ref{eqn:dTdz}) is valid when the number of particles remains fixed -- an excellent approximation after reionization processes.  Table~\ref{table:scalings} lists the different primordial gas heating and cooling processes that contribute additively to the $(3 k_B n_b/2)^{-1} dQ/dt$ term in equation~(\ref{eqn:dTdz}), aside from the column labeled ``Hubble'' which corresponds to the $-2HT$ term.  All of the listed processes are well approximated as power laws in $T_4 \equiv T/10^4$K, and the power-laws in $\Delta$ and $Z_3 \equiv (1+z)/4$ are exact.  The only cooling processes that are not included in Table~\ref{table:scalings} are those due to collisional excitations, which become important at $\Delta \gtrsim 10$ (see \S\ref{ss:collapse}).

Photoheating is the only non-adiabatic heating source included in our calculations and in standard models for the thermal history.  These models appear to be sufficient to describe intergalactic temperature observations over the measured range of $1.5 < z < 5$  \citep{puchwein14, uptonsanderbeck}.  
  Photoheating inputs an energy per \HI\ ionization of
 $\epsilon_{\rm HI} \approx E_{\rm ion, HI}/(2 +\alpha_{\rm bk})$ before \HeII\ reionization (dominated by \HI\ photoheating; $E_{\rm ion, HI}=13.6~$eV), and  $\epsilon_{\rm HI} \approx 3 \, {\rm E_{ion, HI}}/(2 +\alpha_{\rm bk})$ after (dominated by \HI\ and \HeII\ photoheating; see Table \ref{table:scalings}).\footnote{\label{foot:photo} Our \HI\ and \HeI\  photoheating rates assume an abrupt cutoff at $4\;$Ry as found in models \citep[e.g.][]{haardt12}.  For the \HeI\ photoheating rate listed in Table~\ref{table:scalings}, $f(\alpha_{\rm bk})$ is some function of $\alpha_{\rm bk}$ which equals zero for $\alpha_{\rm bk}=0$ and would equal $\alpha_{\rm bk}$ if there were no cut-off in the spectrum at $4~$Ry.  For \HI, there should be a similar functional form that depends on the cutoff at $E_{\rm max} = 4\,$Ry; the $\alpha_{\rm bk}$ scaling included in the table approximates this function with its $E_{\rm max} \rightarrow \infty$ form (which is accurate to $10\%$).}
  Nicely, the photoheating rates, which for a highly photoionized IGM are $\epsilon_X$ times the recombination rate, only depend on the spectral index of the background specific intensity near the ionization edge of the species in question (and not on the amplitude of the background).   We parametrize the background specific intensity as $J_\nu \propto \nu^{-\alpha_{\rm bk}}$  [erg\,s$^{-1}$\,Hz$^{-1}$\,cm$^{-2}$\,sr$^{-1}$].  Ionizing background models predict $\alpha_{\rm bk} \approx 0-1$ over $\approx 1-10~$Ry, with this value being determined by the intrinsic spectrum of the sources and the reprocessing from intergalactic absorptions (e.g., \citealt{haardt12}).  This range in $\alpha_{\rm bk}$ translates to a $30\%$ uncertainty in the post-reionization photoheating rates.  We choose $\alpha_{\rm bk} =0$ for the calculations in this paper.

Figure~\ref{fig:T0gamma} illustrates how $T_0$ and $\gamma-1$ evolve.  Here, $\gamma-1$ is evaluated at $\Delta=1$; we will show that a single power-law holds at all densities only in the limit of late times.  These calculations are initialized with $T_0 = 20,000\,$K and $\gamma-1=0$ at $z=9$, consistent with simulations and analytic calculations of hydrogen reionization \citep{miresc94, trac08, mcquinn-Xray}, and use the heating and cooling rates that apply before \HeII\ was reionized.  The solid curve includes all processes that affect $T_0$, whereas the other curves turn off the specified process(es) to illustrate how they impact the thermal history.
The main source of cooling at $\Delta \sim 1$ is the expansion of the Universe.  Compton cooling and recombination cooling are smaller but never negligible, with Compton cooling being quite important for shaping $T_0$.  

Figure~\ref{fig:T0gammaHeII} is the same as Figure~\ref{fig:T0gamma} except that it considers the case starting with $T_0 = 15,000\,$K and $\gamma-1=0$ at $z=3.5$, values more consistent with the IGM after \HeII\ was reionized \citep{mcquinn09, becker10}, and it takes the applicable rates for after \HeII\ reionization.  In this case, Compton cooling is less important and neglecting non-adiabatic cooling processes is a much better approximation.

 \section{Derivation of $T-\Delta$ relation ignoring non-adiabatic cooling}
\label{sec:asymp}

Our aim is an analytic understanding of why the temperature is driven to a single $T-\Delta$ relation.  Let us introduce the adiabatic variable $\eta = T/n_H^{2/3}$, using that $n_H =\bar n_{H,0}\, a^{-3} \Delta(a)$, where $n_H$ is the total number of hydrogen (\HI\ + \HII), subscript ``$0$'' indicates the present day value, and $\Delta(a)$ is the density contrast with respect to the mean.  The time derivative of $\eta$ is
\begin{equation}
n_H^{2/3} \frac{d\eta}{dt} = \frac{dT}{dt} + 2 H T - \frac{2 \, T}{3 \Delta} \frac{d \Delta}{dt},
\label{eqn:eta}
\end{equation}
such that, after substituting this into equation~(\ref{eqn:dTdz}), our equation for $dT/dt$ becomes
\begin{equation}
n_H^{2/3} \frac{d\eta}{dt} = \frac{2}{3k_Bn_b}\frac{dQ}{dt}.\label{eqn:detadt}
\end{equation}

We first specialize to the simplified case of an IGM with only photoheating and adiabatic processes, the limit considered in \citet{hui97}.  While we saw in \S\ref{ss:processes} that this limit is not a good approximation for the evolution of $T_0$ at high redshifts owing to Compton cooling off of the CMB, this limit is a better approximation for the evolution of $\gamma-1$ (Fig.~\ref{fig:T0gamma}).  In addition, for calculations that start at lower redshifts, this photoheating-only limit is an excellent approximation for both $T_0$ and $\gamma-1$ (Fig.~\ref{fig:T0gammaHeII}).  With only the photoheating contribution to $dQ/dt$, equation~(\ref{eqn:detadt}) becomes
\begin{equation}
n_H^{2/3} \frac{d\eta}{dt} =  \frac{2 \epsilon_{\rm HI} \Gamma_{\rm HI} n_{\rm HI}}{3k_Bn_b}, 
\label{eqn:justphotoequildt}
\end{equation}
where $\Gamma_{\rm HI}$ is the \HI\ photoionization rate.   Assuming that the gas is in photoionization equilibrium so that $n_{\rm HI} \approx \alpha_{\rm A} n_e n_H/\Gamma_{\rm HI}$ and changing the time variable to the scale factor, $a$, yields 
\begin{equation}
n_H^{2/3} \frac{d\eta}{da} =  \frac{\epsilon_{\rm HI} \alpha_{\rm A} n_H a^{1/2}}{3 k_B H_0 \Omega_m^{1/2}}. 
\label{eqn:justphotoequil}
\end{equation}
Note that to reach this equation we assumed $n_e \approx n_H$ and $H = H_0 \sqrt{\Omega_m} a^{-3/2}$, appropriate in the assumed cosmology at $z \gtrsim 1$.  In what follows, we use that the CASE A recombination coefficient can be written as $\alpha_{\rm A}(T) = \alpha_{\rm A}(T_p) \times (T/T_p)^{-0.7} = \alpha_{\rm A}(T_p) \times (n_H^{2/3} \eta/T_p)^{-0.7}$, where $T_p$ is some reference temperature.  Since it makes the algebra cleaner, we will also approximate $-0.7$ as $-2/3$, but will comment on the generalization. Integrating equation~(\ref{eqn:justphotoequil}) over time yields
\begin{eqnarray}
\int_{\eta_i}^{\eta} d\eta \, \eta^{2/3}  &=& A \int_{a_i}^{a} \,da \,a^{1/2}\,n_H^{-(2/3)^2 + 1 - 2/3};\\
A &\equiv & \frac{\alpha_{\rm A}(T_p) T_p^{2/3} \epsilon_{\rm HI}}{3 k_b H_0 \Omega_m^{1/2}},
\end{eqnarray}
or
\begin{equation}
\eta^{5/3} -  \eta_i^{5/3}  = \frac{5}{3} A  \, \bar n_{H,0}^{-1/9} \int_{a_i}^{a} \, da' \, {a'}^{5/6} \Delta(a')^{-1/9}.
\label{eqn:eta_ev}
\end{equation}
Since the density of most intergalactic gas decreases considerably over a cosmologically significant period, whereas the temperature decreases only by as much as a factor of $\sim 2$, $(\eta_i/\eta)^{5/3}$ decreases strongly with increasing time, erasing memory of the initial conditions as $\sim (a_i/a)^{10/3}$ for gas at the mean density.  Taking the limit $\eta \gg \eta_i$ and parametrizing the density evolution of a gas parcel as $\Delta(a') = \Delta(a) \times (a'/a)^{n}$, equation~(\ref{eqn:eta_ev}) becomes
\begin{eqnarray}
T(a)^{5/3}  &=& \frac{5  A \, \bar n_{H,0}  \,a^{-3/2} \Delta(a) }{3\, (11/6 -n/9)}  \left(1- \left[\frac{a_i}{a} \right]^{11/6-n/9} \right).
\label{eqn:Tstar}
\end{eqnarray}

Equations~(\ref{eqn:eta_ev}) and (\ref{eqn:Tstar}) demonstrate that the asymptotic temperature of a gas element depends very weakly on whether it collapses to its current density, $\Delta(a)$, quickly or slowly.  The time dependence of $\Delta(a')$ only enters in the temporal integral on the RHS of equation~(\ref{eqn:eta_ev}) and then to the $-1/9$ power.  
Equation~(\ref{eqn:Tstar}) shows that unless $|n| \gtrsim 9$, the collapse/expansion has less than a factor of two effect on the proportionality constant in the asymptotic $T -\Delta$ relationship relative to the $n=0$ case.  The exponent $n=9$ corresponds to $\Delta(t)$ changing by a factor of two over a timescale of $da/a = 0.1$.  Since $\Delta a/a = H \Delta t $, and the dynamical time is $H^{-1} \Delta^{-1/2}$ for $\Delta \gg 1$, $\Delta a/a \approx 0.1$ corresponds to the dynamical time for gas with $\Delta \approx 100$.  Using these relations, the expected $n$ for $\Delta = 10$ gas (the highest densities of interest) is $\approx 3$, which would lead to a $20\%$ deviation from the $n=0$ normalization.  Within trajectories in the Zeldovich approximation (where the collapse happens too suddenly at high densities), we find visually a range of $5 \lesssim n \lesssim 7$ for $\Delta=10$ and $z=3$,\footnote{This is the effective $n$ that corresponds to the instantaneous evolution of $d\Delta/dt$.} with $n=5$ resulting in $15\%$ smaller temperatures than $n=7$ in equation~(\ref{eqn:Tstar}).  
In detail, the Zeldovich calculations shown in the bottom panel of Figure~\ref{fig:T0gamma} have a standard deviation in $\Delta T/T$ of $5\%$ at $\Delta=10$, and this becomes somewhat smaller at lower $\Delta$.  \emph{Thus, the tightness of the $T-\Delta$ relation owes largely to the weak dependence of the final temperature on the previous collapse or expansion history of a gas element.}  The average dependence of $n$ on $\Delta$ does induce a small excess slope to the $T-\Delta$ relationship that we estimate in \S\ref{sec:other_cool}.

In the $n= 0$ case, equation~(\ref{eqn:Tstar}) becomes 
\begin{eqnarray}
T(a, \Delta) &=& T_{0, \rm lim}(a) \Delta^{3/5} \label{eqn:T0lim}\\
T_{0, \rm lim}(a) &=& \left( \frac{10\alpha_{A}(T_p) T_p^{2/3} \epsilon_{\rm HI} \bar n_{H,0} a^{-3/2} (1- [\frac{a_i}{a}]^{11/6})}{33 k_b H_0 \Omega_m^{1/2}} \right)^{3/5}, \nonumber \\
 & =&  9400 {\rm \, K \,} \left(\frac{\epsilon_{\rm HI} }{5 {\rm eV}}\right)^{3/5} \left(\frac{a \left(1- [\frac{a_i}{a}]^{11/6}\right)^{-2/3} }{0.25}\right)^{-9/10},\nonumber
\end{eqnarray}
although we emphasize that the result changes little for any $n$ that is likely to apply in the low density IGM.   Note again that $\epsilon_{\rm HI} \approx 13.6 {\rm eV}/(2 + \alpha_{\rm bk})$ before \HeII\ was reionized, and it equals three times this value afterwards.

In conclusion, in the photoheating-only limit we find $T \propto \Delta^{3/5}$, in excellent agreement with the low-redshift asymptote of the solution to $T(z, \Delta)$ with all cooling processes included.  (If we had instead used the parametrization $\alpha_{\rm A} \sim T^{-\beta}$ for the recombination rate, we would find $T \propto \Delta^{1/(1+\beta)}$, with $\gamma-1 = 0.59$ for $\beta = 0.7$.)  The physical reason for this asymptotic relation is quite apparent from the form of equation~(\ref{eqn:justphotoequildt}):
\begin{equation}
\frac{d\eta}{dt} = \frac{q_{\rm photo}(\eta, n_H)}{n_H^{2/3}},
\end{equation}
where $q_{\rm photo}(\eta, n_H)$ is the heating rate of the gas.  Because $q_{\rm photo}/{n_H^{2/3}}$ is almost independent on $n_H$ at fixed $\eta$ at $z\lesssim 6$ (scaling as $\eta^{-0.7} n_H^{0.1}$), all gas is driven towards the same adiabat, erasing memory of the initial conditions.  Because the heating is nearly constant along an adiabat, this constancy also implies that the final temperature of a gas parcel depends extremely weakly on its collapse history.  We now show that these conclusions hold when all relevant heating and cooling processes are included.

\section{the relation with all cooling processes included}
\label{sec:other_cool}

So far we have considered the case of photoheating and only adiabatic cooling processes.  Taking $T_4=1$, $\Delta =1$, $Z_3=1$, and $\alpha_{\rm bk}=0$, the amount of atomic and free-free cooling is about $50\%$ of the amount of photoheating before \HeII\ was reionized, and about $20\%$ afterwards (see Table~\ref{table:scalings}).  Larger percentiles occur for other parameter choices.  Thus, it is plausible that cooling processes break the tightness or power-law form of the $T-\Delta$ relation at certain epochs and certain densities.  This section first provides an initial sketch of why cooling processes do not break the relation.  Then, in \S\ref{ss:comp} we treat the case of Compton cooling in more detail.
 
Let us treat the non-adiabatic coolants as a small correction to our previous solution such that $\eta = \eta^{(0)} + \epsilon \eta^{(1)}$, where $\eta^{(0)}$ is the photoheating-only solution and $\eta^{(1)}$ is the perturbation around this solution including all other processes. We also multiply all cooling terms that are not in our ``zeroth order'' equation, equation~(\ref{eqn:justphotoequil}), by $\epsilon$. Collecting the terms that are first order in $\epsilon$ yields the equation
\begin{eqnarray}
n_H^{2/3} \frac{d\eta^{(1)}}{dt} +  0.7 \dot q_{\rm photo}^{(0)} \frac{\eta^{(1)}}{\eta^{(0)}} =  -\left( \dot q_{\rm Comp}^{(0)} +  \dot q_{\rm rec}^{(0)}+  \dot q_{\rm ff}^{(0)} \right),
\label{eqn:second_order}
\end{eqnarray}
where we use the notation that $\dot q_{X}^{(0)}$ is the heating/cooling rate of $X$ [degrees Kelvin per second per particle] evaluated at $\eta^{(0)}$ and $n_H$, and where we have grouped all recombination cooling rates into $\dot q_{\rm rec}$.   To derive a rough estimate for the size of the correction, we further assume that the $d\eta^{(1)}/{dt}$ term is smaller than the others so that this equation becomes
\begin{eqnarray}
0.7 \dot q_{\rm photo}^{(0)} \frac{T^{(1)}}{T^{(0)}}  = -\left( \dot q_{\rm Comp}^{(0)} +  \dot q_{\rm rec}^{(0)} +  \dot q_{\rm ff}^{(0)} \right).
\label{eqn:second_order_equil}
\end{eqnarray}
We find numerically that this equilibrium solution is a decent approximation for the full solution to equation~(\ref{eqn:second_order}).  
Using Table~\ref{table:scalings} to insert physical values for the heating rates, equation~(\ref{eqn:second_order_equil}) becomes
\begin{equation}
 \frac{T^{(1)}}{T^{(0)}} \approx -\overbrace{0.35 Z_3}^{\rm Compton}  - \overbrace{0.15 \Delta^{0.6} \left(\frac{T_0^{(0)}}{10^4 {\rm K}}\right)}^{\rm recombination + ff}
 \label{eqn:second_order_equil_values}
\end{equation}
before the \HeII\ was reionized and for $\alpha = 0$.  To represent the formula in this simple form, we assumed that the recombination cooling and free-free cooling rates scale as $T^{0.3}$, bisecting the scaling of these different rates. If we had instead substituted the applicable rates for after \HeII\ reionization, the fractional contribution of recombination plus free-free cooling would stay nearly the same, while the contribution of Compton cooling would be reduced by a factor of nearly three relative to equation~(\ref{eqn:second_order_equil_values}).

Equation~(\ref{eqn:second_order_equil_values}) shows that Compton cooling does not alter the $T-\Delta$ relation. This occurs simply because $\dot q_{\rm Comp}^{(0)} \propto T^{(0)}$ with no direct $\Delta$ dependence.  Recombination cooling and free-free cooling do alter the $T-\Delta$, resulting in its fractional contribution to $T^{(1)}$ having a strong dependence on $\Delta$, as $\Delta^{0.6}$ in equation~(\ref{eqn:second_order_equil_values}).  However, even at the highest densities where the $T-\Delta$ relation applies, $\Delta \sim 10$, recombination and free-free cooling contribute a significant $40 \%$ correction for $T_{\rm 0, lim} = 7000~$K.  Using Equation~(\ref{eqn:second_order_equil_values}), the correction from recombination and Compton cooling to the equation of state results in a flatter value of
\begin{equation}
\gamma-1 \approx 0.59 - \overbrace{\frac{0.06}{1-0.35 Z_3} \Delta^{0.6}  \left(\frac{T_0^{(0)}}{7000 {\rm K}}\right)}^{\rm recombination + ff} +  \overbrace{\frac{0.03}{1-0.35 Z_3} \Delta^{0.5}}^{\rm avg.~collapse~history},
\label{eqn:gamma}
\end{equation}
assuming $\alpha_{\rm bk} =0$ and using the $\beta=0.7$ value for the leading order rather than our simplification of $\beta=2/3$. (Ignore the ``avg.~collapse~history'' term in this expression momentarily.)  A similar correction to the ``recombination+ff'' term in the above equations is also noted seen in the full calculation of $T(z)$: Compare the low-redshift asymptote of the ``all processes'' curve to the ``no recombination cooling curve'' in the bottom panel of Figure~\ref{fig:T0gamma}.  

The ``avg. collapse history'' term in equation~(\ref{eqn:gamma}) is the correction from adiabatic heating (the $n$-dependent term in $T_0^{\rm lim}$).  In particular, we assumed that this correction scales as $\Delta^{0.5}$ as expected if the characteristic time is the dynamical time, with $n=3$ corresponding to $\Delta=10$ and $n=0$ to $\Delta =1$ as argued for in \S\ref{sec:asymp}.\footnote{We note that the ``avg.~collapse~history'' correction is exactly what is needed to explain the $\gamma=0.62$ asymptote derived using linear theory in \citet{hui97} for the photoheating-only limit.}  This correction acts to cancel half of the  ``recombination+ff'' term.

 The above assumed that Compton cooling can be treated perturbatively, which Figure~\ref{fig:T0gamma} shows is a dubious assumption, especially at high redshifts (although it looks to be a good assumption for the case shown in Fig.~\ref{fig:T0gamma}).  We show in \S\ref{ss:comp} that Compton cooling also does not affect the asymptotic $\gamma-1$ even when this process shapes the thermal history.

\subsection{Full treatment of Compton cooling}
\label{ss:comp}
Here we calculate $T(\Delta, a)$ in the presence of Compton cooling.  Unlike the previous section, we do not make the dubious assumption that Compton cooling contributes a small correction to the full solution.  The evolution of the temperature for the case of photoheating plus Compton cooling can be solved exactly.  The differential equation for the temperature becomes (after a few simple manipulations) 
\begin{equation}
\frac{3}{5} \frac{d\eta^{5/3}}{da} + C a^{-7/2}  \eta^{5/3} =  A a^{1/2} n_{\rm H}^{-(2/3)^2 +1-2/3},
\label{eqn:all}
\end{equation}
where 
\begin{equation}
C= \frac{4 \sigma_T a_{\rm rad} T_{\rm cmb,0}^4 }{3 m_e c H_0 \Omega_m^{1/2} } \approx 0.011,
\end{equation}
and $a_{\rm rad}$ is the radiation constant.
 We solve 
\begin{equation}
\frac{3}{5} \frac{dG}{da} + C a^{-7/2} G = \delta(a - a'),
\end{equation}
for the Green's function $G$ of the linear in $\eta^{5/3}$ part of equation~(\ref{eqn:all}), which yields
\begin{eqnarray}
G(a, a') &=&  \frac{5}{3}  e^{-\frac{2}{3}C({a'}^{-5/2}- a^{-5/2})} \theta(a - a'),
\end{eqnarray}
where $\theta$ is the Heaviside function.  The solution to equation~(\ref{eqn:all}) is the sum of a term that satisfies the linear in $\eta^{5/3}$ part of this equation plus the integral of $G$ over the source term:
\begin{eqnarray}
\eta^{5/3} &=& \eta^{5/3}_i  e^{-\frac{2}{3}C (a_i^{-5/2} - a^{-5/2}) } \label{eqn:withcompton}\\ 
& +&  \frac{5}{3} A n_{H,0}^{-1/9} \int_{a_i}^a da'  {a'}^{5/6}  e^{-\frac{2}{3}C ({a'}^{-5/2} - a^{-5/2})} \Delta(a')^{-1/9}. \nonumber
\end{eqnarray}
Notice that our equation is identical to equation~(\ref{eqn:eta_ev}) in the limit of no Compton cooling ($C\rightarrow 0$).  Our solution also reduces to the solution for $T_0$ in Appendix A of \citet{lidz15} for the case $\Delta(a) = 1$.  At high redshifts, Compton cooling enhances the rate that memory of the initial conditions is erased via the factor $e^{-2/3 C (a_i^{-5/2} - a^{-5/2})}$.  Equation~(\ref{eqn:withcompton}) also shows that once the memory of the initial conditions is erased, $\eta^{5/3} \propto \Delta^{-1/9}$ or equivalently $T \propto \Delta^{3/5}$, the same asymptote that we found in the photoheating-only limit.  Also as in the photoheating-only limit, the history of the density contrast enters in the temporal integral to the $-1/9$ power, again making the result insensitive to the temporal evolution of $\Delta(a)$.

We find that Compton cooling results in an extra factor in the asymptotic relation for $T(\Delta)$ over our photoheating-only solution (eqn.~\ref{eqn:T0lim}):
\begin{eqnarray}
T(a, \Delta) &=& T_{\rm 0, lim}^{\rm Comp}(a)~ \Delta^{3/5},  {\rm ~~~~~where} \nonumber\\
\frac{T_{\rm 0, lim}^{\rm Comp}}{T_{\rm 0, lim} } &=&  \overbrace{\left(\frac{ {\Gamma}\left[-\frac{11}{15}, \left(\frac{a_{\rm crit}}{a} \right)^{5/2}\right] - {\Gamma}\left[-\frac{11}{15}, \left(\frac{a_{\rm crit}}{a_i} \right)^{5/2}\right]}{15/11  e^{-\left(\frac{a_{\rm crit}}{a} \right)^{5/2}} a_{\rm crit}^{-11/6}  \left({a}^{11/6} - {a_i}^{11/6} \right)}\right)^{3/5}}^{X_{\rm Comp}}, \nonumber
\end{eqnarray}
where $\Gamma$ is the incomplete gamma function, $a_{\rm crit}  = [2 C/3]^{2/5} = 0.14$, and we have assumed $n=0$.  Evaluating $X_{\rm Comp}$ numerically yields $\{0.81,~ 0.78,~0.76\}$ at $z=\{3, 4, 5\}$ for $z_i = 9$.  For $z_i=6$, $X_{\rm Comp}$ instead yields $\{0.88,~0.89, 0.92\}$.  
  These small reductions in the asymptotic $T_0$ are not the reason Compton cooling has such a large effect in Figure~2;  the primary effect of Compton cooling is in its exponential suppression for $a> a_{\rm crit}$ of $T_i$.

In conclusion, while the asymptotic $T_0$ is marginally altered by Compton cooling off of the CMB, the asymptotic $\gamma-1$ is unaffected by it.   Compton cooling does not increase the sensitivity to a gas parcel's collapse/expansion history, and it accelerates the convergence to the asymptotic $T(\Delta)$.

\section{applications of results}
\label{sec:applications}

Here we use our results to understand (1) how quickly temperature fluctuations are erased (\S\ref{ss:asymp}), (2) the importance of shock heating (\S\ref{sec:shocks}), and (3) the thermal evolution of collapsing gas clouds (\S\ref{ss:collapse}).

\subsection{timescale to reach the asymptote and erase temperature fluctuations}
\label{ss:asymp}

Let us consider a toy universe that has an isothermal $T-\Delta$ relation after reionization.  The temperature at a later time follows from equation~(\ref{eqn:withcompton}):
\begin{eqnarray}
T^{5/3} &=& [T_{\rm 0, lim}^{\rm Comp}]^{5/3} \Delta +  \left( \frac{a^{-3} \Delta}{a_i^{-3}\Delta_i} \right)^{10/9}  {\cal C} \, T_i^{5/3},\\
&& {\cal C}  \equiv e^{- a_{\rm crit}^{5/2} (a_i^{-5/2} - a^{-5/2})},
\end{eqnarray}
 where $z_{\rm crit} = 1/a_{\rm crit} - 1=6.15$ and $T_i$ is the temperature after reionization.  The timescale for the $T_i$--dependent term to become a faction $f$ of $[T_{\rm 0, lim}^{\rm Comp}]^{5/3}$ is 
\begin{equation}
a/a_i \approx (f \Delta_i )^{-1/3}  {\cal C}^{1/3} \, [T_i/ T_{\rm 0, lim}^{\rm Comp}]^{5/9},
\label{eqn:Trelax}
\end{equation}
 where we have approximated the $10/9$ exponent as unity.  For example, if $T_i$ is twice $T_{\rm 0, lim}^{\rm Comp}$ -- roughly the multiple expected after \HI\ and \HeII\ reionization --, it takes $a/a_i  \approx 1.9 ({\cal C}/\Delta_i)^{1/3}$ for the temperature to equal half of its limiting value ($f=0.5$).  Interestingly, this timescale is slightly shorter in initially overdense regions and there is no dependence on the final density.

Fluctuations in temperature -- an inevitable byproduct of a temporally extended and inhomogeneous reionization process -- redshift away in a nearly identical manner to deviations from the asymptotic temperature.  After reionization, the spatial variance of $\eta^{5/3}$ is simply
\begin{equation}
{\rm var} [\eta^{5/3}]_\Delta  =   {\cal C}^2 {\rm var} [ \eta_i^{5/3}]_\Delta,
\end{equation}
where ${\rm var}[Y]_\Delta$ is the variance of $Y$ fixing the final density $\Delta$ but averaging over all spatial positions, and the subscript $i$ again denotes some initial time.  The fractional variance in the temperature in terms of the mean is then
\begin{equation}
{\rm var} \left[\left(\frac{T}{\langle T \rangle}\right) \right]_{\Delta} \approx {\cal C}^2 \, {\rm var}\left[ \frac{ \eta_i}{\langle{\eta} \rangle}  \right]_{\Delta},
\end{equation}
specializing to times when $\Delta T/T \ll 1$ so that ${\rm var}[Y^m] \approx m^2 {\rm var}[Y]$.  This approximation likely holds soon after reionization processes as simulations show that $\Delta T/T \sim 1$ at the end of \HI\ and \HeII\ reionization \citep{trac08, mcquinn09, compostella13}.  Thus, the RMS of temperature fluctuations decreases over essentially the same timescale the average temperature asymptotes, again with a slightly quicker relaxation for initially denser regions. 

The above discussion relates the temperature fluctuations generated at some ``initial'' time to those at some time thereafter.
A better way to conceptualize how temperature fluctuations are imprinted, particularly during hydrogen reionization, is that a region is heated to a similar temperature when it is reionized but different regions are reionized at different times.  The duration of reionization and the associated cooling results in a spread in temperature \citep{trac08, furlanetto09}.  In this scenario, the above relations can be used to calculate how the temperature fluctuations relate to the duration of reionization.

\subsection{the importance of shocks}
\label{sec:shocks}

\begin{figure}
\begin{center}
\epsfig{file=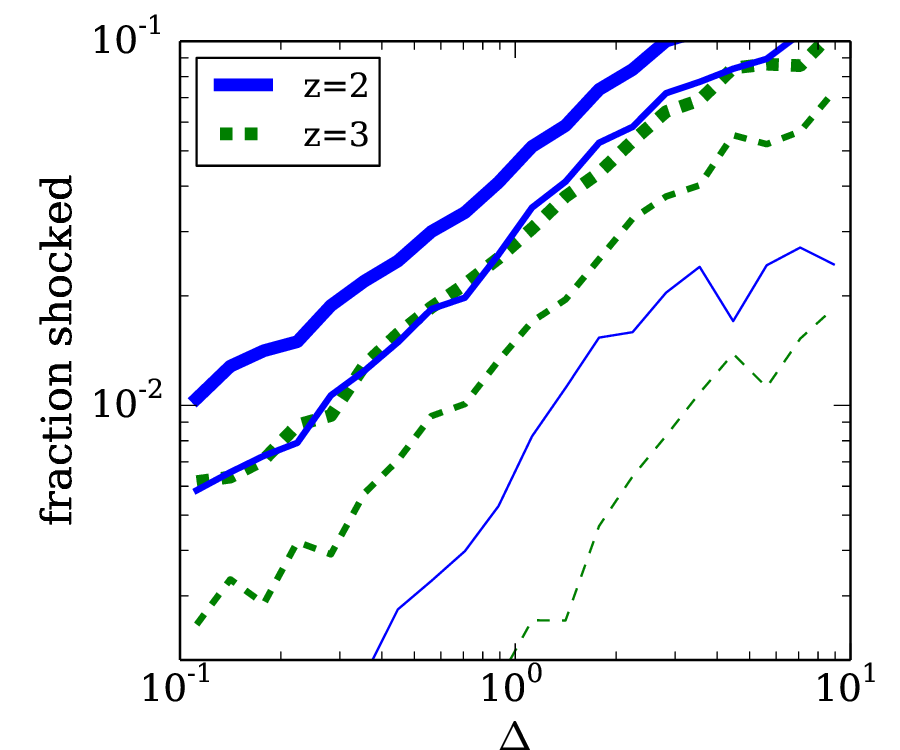, width=8cm}
\end{center}
\caption{Fraction of gas elements in our cosmological simulation that are shock heated by a factor of $1.5, ~2$ and $10$ over the mean $T(\Delta)$ (thick, medium width, and thin curves, respectively) at $z=2.3$ (solid curves) and $z=3.0$ (dashed curves).  The analytic calculations in this paper are valid to the extent shock heating is not important.
\label{fig:shocking}}
\end{figure}

We have shown that without shocks a very tight $T-\Delta$ relation should hold at $\Delta \lesssim 10$ (see Fig.~1).  Thus, in a cosmological simulation, any deviations of the fluid elements from a tight $T-\Delta$ relation indicates shocking.  The black points in Figure~1 show the $T(\Delta)$ of $300$ particles randomly selected from a cosmological simulation.  The particles do not show as tight a $T-\Delta$ relation as in the Zeldovich approximation calculations (red points), suggesting that some of the gas elements in the simulation have experienced weak shocking, although most particles fall within $10\%$ of the expected relation.  Figure~\ref{fig:shocking} shows the fraction of gas that is heated by a factor of $1.5, ~2$ and $10$ over the mean $T(\Delta)$ as a function of $\Delta$ (thick, medium width, and thin curves, respectively) at $z=2.3$ and $z=3$.  By $z=2.3$, the simulation shows that only a few percent of fluid elements with $\Delta = 1$ have been shock heated significantly, with this fraction scaling as $\sim \Delta^{0.6}$.  The shocked fraction in the simulation decreases quickly with increasing redshift.  

\subsection{the thermal history of fluid elements}
\label{ss:collapse}
\begin{figure}
\begin{center}
\epsfig{file=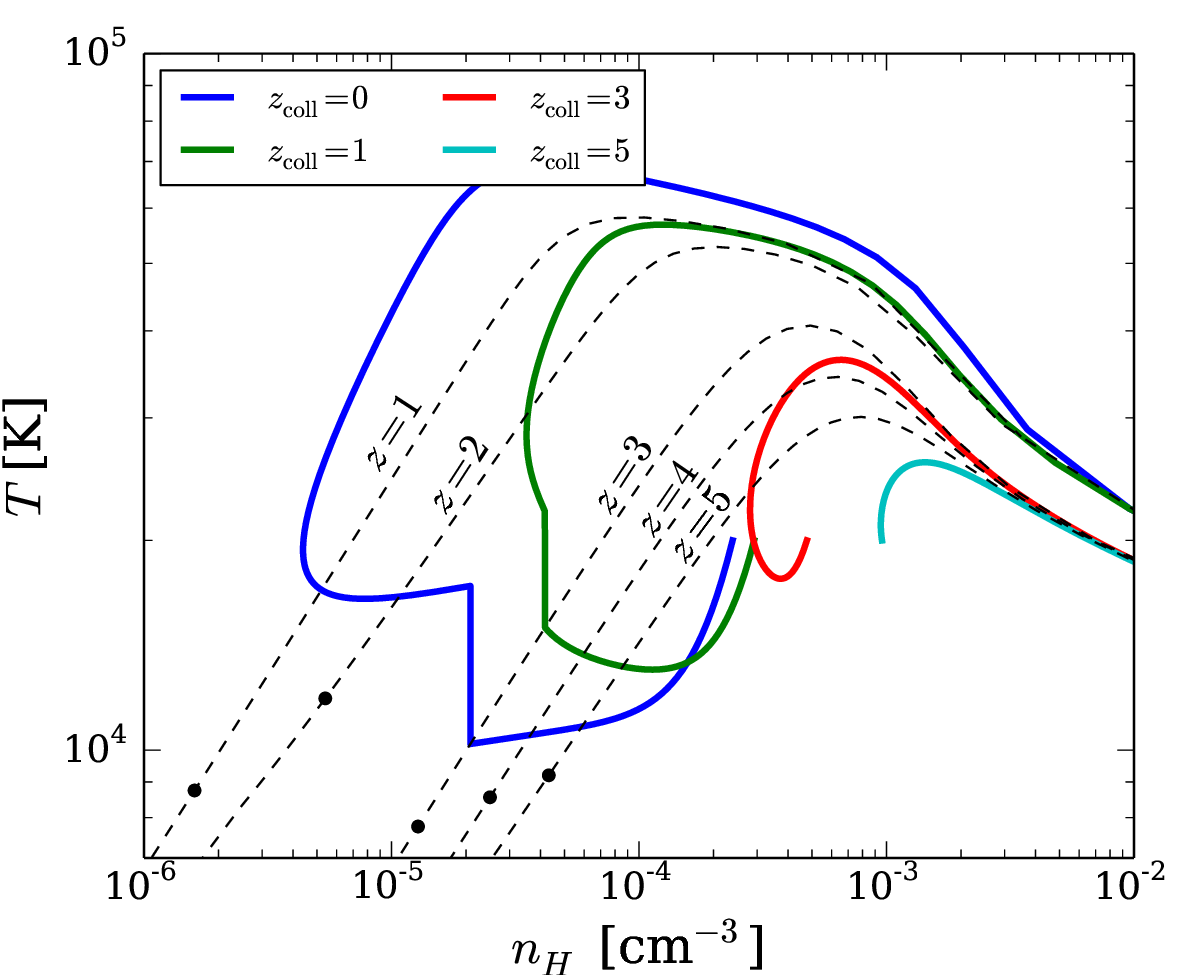, width=8cm}
\end{center}
\caption{Solid curves show the $T(\Delta)$ trajectories of gas parcels that collapse at the specified redshifts, assuming spherical top hat collapse and assuming that the gas was heated by reionization to $20,000\;$K at $z=9$.  Gas parcels start at this temperature and evolve with time in a clockwise fashion in this plot.   The dashed curves show the $T-\Delta$ relation at the specified redshifts, with the filled dot on each curve indicating the temperature at the cosmic mean density.  These curves show that a power-law relation generally holds about a decade in density above the cosmic mean, after which collisional cooling processes become important.  These calculations also assume a heat injection of $7000\;$K at $z=3$ to approximate the heating from \HeII\ reionization.
\label{fig:Tev}}
\end{figure}

The maximum Jeans' mass of a collapsing matter cloud, and whether this mass is larger than the cloud's gravitational mass, determines whether it can overcome pressure and collapse on to a galaxy \citep[e.g.][]{noh14}.  Since the Jeans' mass depends on the gas temperature, the thermal history of collapsing fluid elements sets the minimum mass of galaxies.  The solid curves in Figure~\ref{fig:Tev} show the thermal history of unenriched gas that collapses at the specified redshifts.  These curves are calculated assuming spherical top hat collapse, that the gas was heated to $20,000\;$K at $z=9$ by the reionization of \HI\ and \HeI, that it was kept ionized with a background of $\alpha_{\rm bk} \approx 0$ thereafter, and that there was an instantaneous reionization of \HeII\ at $z=3$ (and an associated heat injection of $7000~$K).  While modeling reionization processes as instantaneous is artificial, the thermal history of this scenario roughly approximates the measured history \citep{becker10}.  The trajectory of a fluid element is clockwise in this plot, with $n_H \rightarrow \infty$ as $z \rightarrow z_{\rm coll}^+$.\footnote{Of minor importance to our results, we have also assumed that at densities where collisional cooling processes become important, the \HeII\ is self-shielding to ionizing backgrounds, as motivated in \citet{noh14}, so that \HeII\ collisional cooling can occur.}

The dashed curves in Figure~\ref{fig:Tev} show the $T-\Delta$ relation at the specified redshifts, with the filled dot on each curve indicating the temperature at the cosmic mean density.  These curves can be used as a clock to infer the redshifts of different points along the (solid) collapsing curves.  They also show that a power-law $T-\Delta$ relation generally holds about a decade in density above the cosmic mean, to densities at which collisional cooling processes become important.\footnote{Perhaps surprisingly, the asymptote of the solid curves bifurcates into two values as $n_H \rightarrow \infty$.  We believe this owes to the two peaks in the cooling function of primordial gas as a function of temperature (owing respectively to \HI\ and \HeII\ cooling), resulting in two thermally stable equilibria.  There is evidence that a similar bifurcation happens in cosmological simulations \citep{noh14}.}  The solid curves in Figure~\ref{fig:Tev} resemble the qualitative picture for a collapsing fluid element's trajectory in $T-\Delta$ space presented in \citet{noh14}, which this study then used to infer the minimum mass of galaxies. 

\section{previous explanations for $\gamma$}
\label{sec:previous_explanations}
We are aware of two previous attempts to provide an intuitive explanation for the $T-\Delta$ relation.  First, \citet[][Appendix C]{theuns98} explained the relation as owing to the balance between photoheating and adiabatic cooling from the uniform Hubble expansion.  \citet{theuns98} derived the equation $\gamma-1 = 1/(\beta+1)$ with $\beta \approx 0.7$, the same relation derived in \S\ref{sec:asymp}.  However, we think this approach gives the correct result from an incorrect insight, as the asymptotic $\gamma-1$ has nothing to with cosmological expansion as this approach posits.  Our derivation shows that photoheated gas that cannot cool converges to the same $\gamma-1$, virtually independent of its collapse or expansion history.  We offer more quantitative musings about the \citet{theuns98} approach in the ensuing footnote.\footnote{The \citet{theuns98} approach is equivalent to taking the photoheating-only temperature equation (eqn.~\ref{eqn:eta}) and dropping the adiabatic expansion/contraction term owing to inhomogeneities (the $2^{\rm nd}$ term on the RHS of eqn.~\ref{eqn:eta}) as well as $dT/dt$ term, such that the resulting formula equates the photoheating rate with the $2HT$ adiabatic cooling rate.  To investigate the \citet{theuns98} approach in detail, let us drop \emph{just} the adiabatic expansion/contraction term owing to inhomogeneities.  The resulting equation is
\begin{equation}
\bar n_H^{2/3} \frac{d\bar \eta}{da} =  \frac{2 \epsilon_{\rm HI} \Gamma n_{\rm HI} a^{1/2}}{3k_Bn_b H_0 \Omega_m^{1/2}}, 
\label{eqn:theuns}
\end{equation}
where bars denote the quantity evaluated at $\Delta=1$.
Solving this equation and taking the limit $\bar \eta \gg \bar \eta_i$ yields a much flatter $\gamma -1 = 0.3$ for collapse histories that have the same $n$.  Figure~\ref{fig:Tdelta} corroborates this result (see the $d\Delta/dt = 0$ curve).  Equation~(\ref{eqn:theuns}) encompasses the one solved in \citet{theuns98}, being more general because we have not imposed $dT/dt=0$.  The approximation of dropping the $d\Delta/dt$ terms is also made in the \HeII\ reionization simulations of \citet{compostella13}, which may explain why their simulations resulted in somewhat smaller $\gamma-1$ than the simulations of \citet{mcquinn09}.}

\citet{puchwein14} provided a separate explanation for $\gamma-1$.  They derived $\gamma-1$ by an argument that is equivalent to taking the $t=\infty$ solution to equation~(\ref{eqn:eta}) for the case of constant $\Delta(t)$.  Thus, they solved $dT/dt = (3/2 n_b k_b)^{-1} [dQ/dt]_{\rm photo}$ in the limit of large times.  The solution to this equation yields the correct slope of $\gamma-1 = 1/(1+\beta)$.  The \citet{puchwein14} approach does hold insight into what sets the asymptotic slope, and indeed is our solution specializing to the case in which the density of a gas parcel is constant in time (our solutions for $n=3$).  However, this explanation ignores the adiabatic cooling associated with the dilation of space, the primary reason why memory of the initial conditions is lost within a doubling time of the scale factor.

A final (and we believe widely held) misconception is that the slope owes to adiabatic processes.  This misconception arises because the asymptotic $\gamma-1$ is close to the expectation of purely adiabatic evolution, $\gamma-1=2/3$.  However, intergalactic gas is constantly ascending to a new adiabat because of photoheating rather than staying on a single adiabat.   This ascension is easily noted in the photoheating-only case where $T_{0, \rm lim} \propto a^{-0.9}$.  This temporal scaling is much different than the scaling $a^{-2}$ that would occur if the evolution were purely adiabatic.  After \HI\ reionization but before \HeII\ reionization, the heating rate per baryon is $4000  (2+\alpha_{\rm bk})^{-1} Z_3 \,\Delta^{0.6}~$K~Gyr$^{-1}$, and it becomes  $12,000 (2+\alpha_{\rm bk})^{-1} Z_3 \,\Delta^{0.6}~$K~Gyr$^{-1}$ after \HeII\ reionization (Table~\ref{table:scalings}).

\section{conclusions}

Many studies have noted that the IGM thermal state asymptotes quickly to a tight power-law $T-\Delta$ relation.  Here we showed that this occurs because the photoheating rate is nearly constant per $n_H^{2/3}$, driving all cosmic gas to nearly a single adiabat in less than a doubling time of the scale factor.  Both cosmic expansion and Compton cooling off of the CMB act to accelerate this convergence.  That a power-law $T-\Delta$ relation arises is a coincidence of how the recombination rate scales with temperature.  We showed that the final temperature of a gas parcel has almost no sensitivity to its previous density evolution, explaining why this relation holds so tightly over two decades in density.  Furthermore, we showed that all relevant cooling processes either reinforce the $T-\Delta$ relation established by photoheating or are just weak enough not to break the relation at relevant densities.  

We were able to derive the full solution to the temperature in the applicable limit of only cosmic expansion and Compton cooling:
\begin{equation}
T= \left( \left[\frac{a_i^3 \Delta}{a^3 \Delta_{i}}\right]^{2/3\times1.7} T_i^{1.7} e^{\left(\frac{0.14}{a} \right)^{5/2} - \left(\frac{0.14}{a_i} \right)^{5/2} }  + [T_{\rm 0, lim}^{\rm Comp}]^{1.7} \Delta \right)^{1/1.7},
\nonumber
\end{equation}
where $T_{\rm 0, lim}^{\rm Comp}$ has an extremely weak dependence on the prior density evolution of a gas parcel.  Further, we showed that other cooling processes cause this solution to deviate at $\lesssim 10\%$ at $\Delta \lesssim 10$, and so this expression can often be used in lieu of a full numerical calculation.

 Cosmological hydrodynamical simulations that employ uniform radiation backgrounds show dispersion in the $T-\Delta$ relation in excess of simple models based on the Zeldovich approximation.  This dispersion owes to shocking (and not to the collapse history of gas clouds).  Such simulations do not capture the fluctuations in the gas temperature that are an inevitable byproduct of reionization processes.  We showed that this additional source of fluctuations is erased over the same timescale that it takes the mean temperature to approach its asymptote.\\

\noindent We thank Joop Schaye for a conversation that motivated this work, and Nick Gnedin, Andrei Mesinger, Adam Lidz, Martin White, and Jennifer Yeh for comments on the manuscript.  MM acknowledges support from NSF award AST 1312724 and HST award HST-AR-13903.001.  This research was supported in part by
the National Science Foundation through XSEDE resources
provided by the San Diego Supercomputing Center
(SDSC) and through award number TG-AST120066.

\bibliographystyle{mn2e}

\bibliography{References}

\begin{thebibliography}{}

\bibitem[\protect\citeauthoryear{{Becker}, {Bolton}, {Haehnelt} \&
  {Sargent}}{{Becker} et~al.}{2011}]{becker10}
{Becker} G.~D.,  {Bolton} J.~S.,  {Haehnelt} M.~G.,    {Sargent} W.~L.~W.,
  2011, \mnras, 410, 1096

\bibitem[\protect\citeauthoryear{{Compostella}, {Cantalupo} \&
  {Porciani}}{{Compostella} et~al.}{2013}]{compostella13}
{Compostella} M.,  {Cantalupo} S.,    {Porciani} C.,  2013, \mnras, 435, 3169

\bibitem[\protect\citeauthoryear{{Furlanetto} \& {Oh}}{{Furlanetto} \&
  {Oh}}{2009}]{furlanetto09}
{Furlanetto} S.~R.,  {Oh} S.~P.,  2009, \apj, 701, 94

\bibitem[\protect\citeauthoryear{{Haardt} \& {Madau}}{{Haardt} \&
  {Madau}}{2012}]{haardt12}
{Haardt} F.,  {Madau} P.,  2012, \apj, 746, 125

\bibitem[\protect\citeauthoryear{{Hui} \& {Gnedin}}{{Hui} \&
  {Gnedin}}{1997}]{hui97}
{Hui} L.,  {Gnedin} N.~Y.,  1997, \mnras, 292, 27

\bibitem[\protect\citeauthoryear{{Lidz}, {Faucher-Gigu{\`e}re}, {Dall'Aglio},
  {McQuinn}, {Fechner}, {Zaldarriaga}, {Hernquist} \& {Dutta}}{{Lidz}
  et~al.}{2010}]{lidz09}
{Lidz} A.,  {Faucher-Gigu{\`e}re} C.,  {Dall'Aglio} A.,  {McQuinn} M.,
  {Fechner} C.,  {Zaldarriaga} M.,  {Hernquist} L.,    {Dutta} S.,  2010, \apj,
  718, 199

\bibitem[\protect\citeauthoryear{{Malloy} \& {Lidz}}{{Malloy} \&
  {Lidz}}{2015}]{lidz15}
{Malloy} M.,  {Lidz} A.,  2015, \apj, 799, 179

\bibitem[\protect\citeauthoryear{{McDonald} et~al.,}{{McDonald}
  et~al.}{2005}]{mcdonald05b}
{McDonald} P.,  et~al., 2005, \apj, 635, 761

\bibitem[\protect\citeauthoryear{{McQuinn}}{{McQuinn}}{2012}]{mcquinn-Xray}
{McQuinn} M.,  2012, \mnras, 426, 1349

\bibitem[\protect\citeauthoryear{{McQuinn}, {Lidz}, {Zaldarriaga}, {Hernquist},
  {Hopkins}, {Dutta} \& {Faucher-Gigu{\`e}re}}{{McQuinn}
  et~al.}{2009}]{mcquinn09}
{McQuinn} M.,  {Lidz} A.,  {Zaldarriaga} M.,  {Hernquist} L.,  {Hopkins} P.~F.,
   {Dutta} S.,    {Faucher-Gigu{\`e}re} C.,  2009, \apj, 694, 842

\bibitem[\protect\citeauthoryear{{Miralda-Escud{\' e}} \&
  {Rees}}{{Miralda-Escud{\' e}} \& {Rees}}{1994}]{miresc94}
{Miralda-Escud{\' e}} J.,  {Rees} M.~J.,  1994, \mnras, 266, 343

\bibitem[\protect\citeauthoryear{{Noh} \& {McQuinn}}{{Noh} \&
  {McQuinn}}{2014}]{noh14}
{Noh} Y.,  {McQuinn} M.,  2014, \mnras, 444, 503

\bibitem[\protect\citeauthoryear{{Puchwein}, {Bolton}, {Haehnelt}, {Madau} \&
  {Becker}}{{Puchwein} et~al.}{2014}]{puchwein14}
{Puchwein} E.,  {Bolton} J.~S.,  {Haehnelt} M.~G.,  {Madau} P.,    {Becker}
  G.~D.,  2014, ArXiv:1410.1531

\bibitem[\protect\citeauthoryear{{Springel}}{{Springel}}{2005}]{springel05}
{Springel} V.,  2005, \mnras, 364, 1105

\bibitem[\protect\citeauthoryear{{Theuns}, {Leonard}, {Efstathiou}, {Pearce} \&
  {Thomas}}{{Theuns} et~al.}{1998}]{theuns98}
{Theuns} T.,  {Leonard} A.,  {Efstathiou} G.,  {Pearce} F.~R.,    {Thomas}
  P.~A.,  1998, \mnras, 301, 478

\bibitem[\protect\citeauthoryear{{Trac}, {Cen} \& {Loeb}}{{Trac}
  et~al.}{2008}]{trac08}
{Trac} H.,  {Cen} R.,    {Loeb} A.,  2008, \apjl, 689, L81

\bibitem[\protect\citeauthoryear{{Upton Sanderbeck}, {D'Aloisio} \&
  {McQuinn}}{{Upton Sanderbeck} et~al.}{2015}]{uptonsanderbeck}
{Upton Sanderbeck} P.~R.,  {D'Aloisio} A.,    {McQuinn} M.~J.,  2015,
  ArXiv:1511.05992

\bibitem[\protect\citeauthoryear{Viel, Haehnelt \& Lewis}{Viel
  et~al.}{2006}]{viel06}
Viel M.,  Haehnelt M.~G.,    Lewis A.,  2006, Mon. Not. Roy. Astron. Soc.
  Lett., 370, L51

\end{thebibliography}

\end{document}